# Diffusion-Dependent Pattern Formation on Crystal Surfaces

Marta Anna Chabowska* and Magdalena A. Załuska-Kotur

Read Online

ACCESS | 📊 Metrics & More | 📄 Article Recommendations

**ABSTRACT:** The growth of a crystal is usually determined by its surface. Many factors influence the growth dynamics. Energy barriers associated with the presence of steps most often decide the emerging pattern. The height and type of Ehrlich–Schwoebel step barriers lead to the growth of nanocolumns, nanowires (NWs), pyramids, and bunches or meanders in the same system. Surface diffusion is another factor that determines the nature of growth. We used the (2 + 1)D cellular automaton model to investigate the additional effect of diffusion along with step barriers. We show that when we change only the diffusion rate, the length of the meanders or the height of the bunches increases, the cracked structure of the nanopillars changes into very long, tall NWs. We show that the length of the step–step correlation is a good characterization of the resulting patterns.

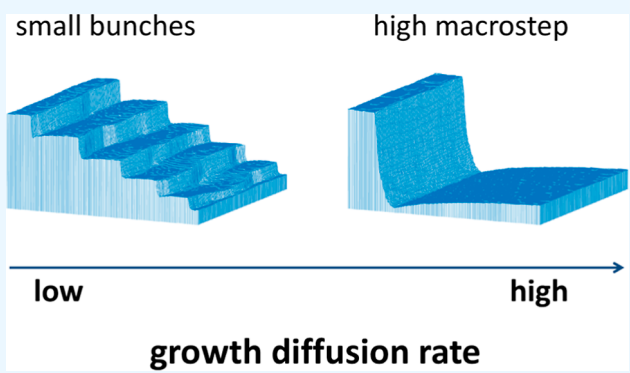

small bunches       high macrostep

low         high

**growth diffusion rate**

## ■ INTRODUCTION

In the era of widespread miniaturization of nanotechnology, the characterization of the crystal surface and, thus, its analysis are more important than ever. Such research turns out to be crucial for conducting the experiments and understanding the obtained results. For example, to understand why nanowires (NWs) only grow on a substrate with certain characteristic,[1] Kang et al. has shown experimentally and theoretically that gold droplets begin to nucleate and guide the growth of NWs only when the {111}B facets become large and regular enough. The results obtained from Monte Carlo simulations show that to maintain the supersaturation conditions in the Au droplet that can initiate NW growth, a minimum step-free collection area is needed as steps inhibit their growth. The authors concluded that the surface morphology of the substrate, including the {111} facet structure, plays a key role in the NW nucleation process. Knowing the surface also allows better control of the growth of structures with the desired geometry and properties, as shown in ref 2, where vicinal surfaces have been used as nanotemplates for the growth low-dimensional systems. In order to have regular planar NW arrays, the authors used highly regular periodic, step-bunched surfaces of n-type doped vicinal Si(111).

From our point of view, the most interesting is the analysis of the vicinal surface, which is the subject of several theoretical and experimental studies.[3−7] This type of surface is important in catalysis during the growth of nanostructures or engineering since steps are active sites for nucleation or chemical reactions.

Many previous investigations have focused on the growth of crystals in different aspects,[8−13] step bunching or meandering.[14−17] In particular, the emergence of the step-bunching instability has been systematically studied.[14] Using the vicinal cellular automaton (vicCA) model, Krzyżewski et al. examined the stability of bunches for the growth and sublimation of 1D vicinal surface in two destabilization modes: step-down and step-up currents. Their detailed analysis was carried out depending on various parameters, such as adatom concentration, diffusion rate, or length of time evolution. They showed that it is possible to reproduce step-bunching instability caused by two opposite drift directions in the two situations of step motion mediating sublimation and growth. In comparison with the (1 + 1)D vicCA model, two-dimensional model (2 + 1)D vicCA is much more realistic and gives chance to get much more different types of structure orderings. We have shown that among the many possible parameters controlling the simulation process, playing only with the presence and height of the direct and inverse Ehrlich–Schwoebel (iES) barriers and the proper selection of the well potential in between lead to the growth of nanocolumns, NWs, and nanopyramids or meanders in the same system.[18]

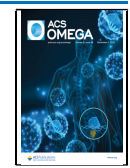

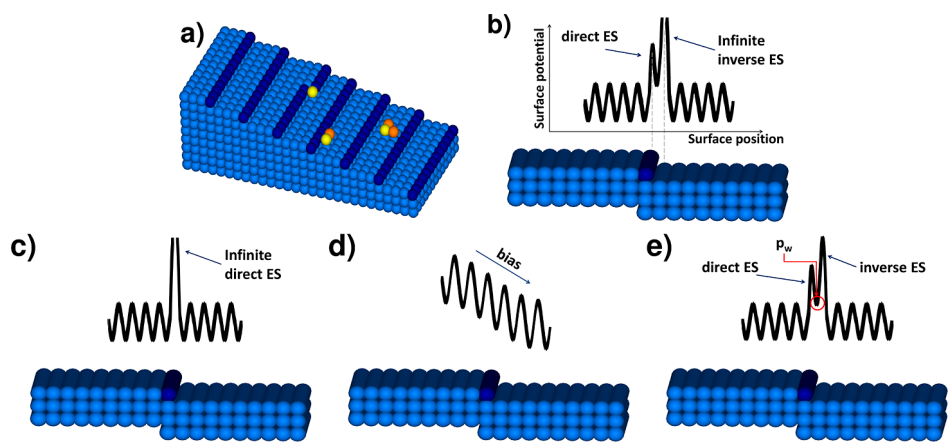

**Figure 1.** Initial conditions of simulations: (a) surface with descending steps from the left to the right and separated by terraces with three situations when an adatom (yellow ball) becomes part of the crystal and visualization of the Ehrlich–Schwoebel (ES) barriers in case of (b) the presence of the infinite inverse ES barrier and high direct dES barrier; (c) lack of the inverse ES and infinite direct ES barrier; (d) lack of both ES barriers and bias; and (e) presence of the inverse ES and direct ES barrier.

Although the changes in the potential energy landscape provide great opportunities to influence the surface pattern formation, it is experimentally difficult to control the potential energy at the surface. It is different in the case of diffusion; we can easily increase its speed by increasing the temperature. With typical diffusion barriers, an increase in temperature of 80 K causes diffusion to accelerate by a factor of approximately 50, approximately. Such temperature changes are easily achieved experimentally. In the present paper, we focus on the 2D vicinal surface and show that in the formation of various structures on it, not only is the combination of step barriers is crucial. We present an analysis of the effect of the diffusion process rate on surface patterns, including the change in surface structure characteristics. We also show the first attempts to compare the obtained structures not only qualitatively but also quantitatively.

## ■ MODEL

The model which we use in this work is (2 + 1)D vicinal cellular automaton model, introduced and studied before in various (1 + 1)D contexts[8,14–17] and (2 + 1)D context.[18] It is built as a combination of two essentially different modules: the cellular automaton (CA) module responsible for the evolution of the vicinal crystal surface and the Monte Carlo (MC) module representing the diffusion of the adatoms. The CA module realizes the growth of the surface on a square lattice according to predefined rules in a parallel fashion while MC module realizes the diffusion of the adatoms in the serial mode, adatom after adatom.

One diffusional step is completed when each adatom is visited once (on average). A single time step of the simulation is represented by the diffusion of all adatoms along the surface (MC unit), then one growth update (CA unit), and finally compensating the adatoms to their initial concentration $c_0$. This design allows the study of large systems in long simulations. Between two growth modules, all adatoms perform diffusional jumps in a serial manner, and their number is denoted by $n_{DS}$, but only jumps that point to a neighboring unoccupied lattice site are made. The diffusional updates do not contribute to increase of the time. Each diffusional jump of particles happens with probability $P$, dependent on the height of the energy barrier to jump over $E_B$ and temperature factor $\beta = \frac{1}{k_B T}$. Jump probability is given by

$$P = e^{-\beta(E_B - E_0)} \tag{1}$$

where we relate jump probability above barrier to the fastest jump process in the system that happens over barrier $E_0$. Below we use symbol $P_{dES}$ for jump over the direct Ehrlich–Schwoebel (ES) barrier and $P_{iES}$ for jumps over inverse ES barrier, which will be explained below. According to eq 1, these probabilities can be changed by changing the temperature of the system. Moreover $n_{DS}$ is a parameter that describes diffusion rate and can be related to temperature by dependence

$$n_{DS} \sim e^{-\beta E_0} \tag{2}$$

Note that $n_{DS}$ is an integer number, so only certain configurations of $\beta E_0$ can be expressed using its value. From eq 2, it is seen that we can control diffusion by temperature. Of course the exact values of temperature increase responsible for changing $n_{DS}$ form 1 to 30 depend on the diffusion barrier $E_0$. For example, if we assume that the barrier is 0.5 eV, this increases temperature to 373 K if we start from 300 K. If $E_0$ is higher, assume that this barrier is 0.8 eV, which is also a possible value in experimental systems, we need to increase temperature to 343 K to realize the choice $n_{DS} = 30$. Other, realizable activation energies and initial temperatures lead to the similar, physically feasible temperature changes. As $n_{DS}$ grows, the system dynamics goes from diffusion-limited (DL) growth toward kinetic-limited (KL) growth mode, and at the same time, the transparency of the step increases.

The model consists of two parts: the 2D surface of the crystal represented by a table with the height of the crystal, given by the number of built-in atom layers, and the second part, the 2D layer above the crystal surface in which randomly distributed adatoms diffuse, allowing to feel an additional one dimension above the surface. Hence, the name of the model is (2 + 1)D. The crystal surface usually consists of descending steps. They fall from left to right and are initially separated by $l_0$ length terraces. In the direction along the steps, periodic boundary conditions are imposed, while across the steps, helical periodic boundary conditions are applied to maintain the step differences.

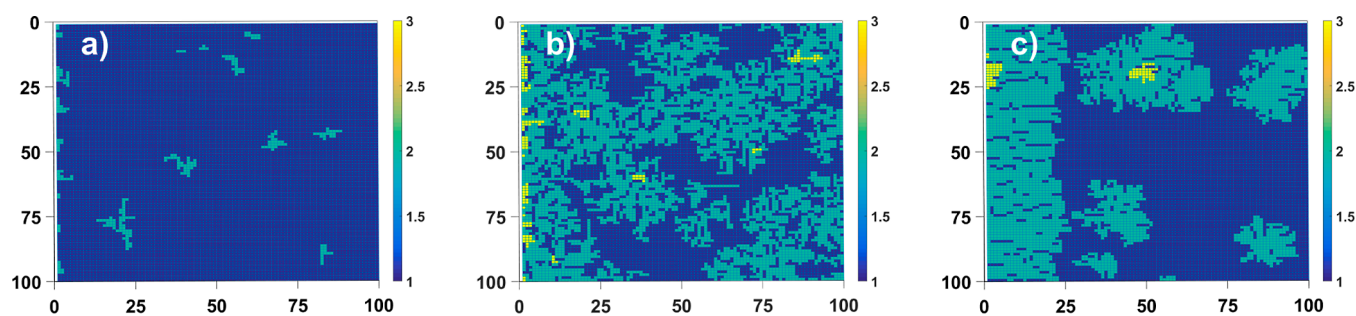

**Figure 2.** Islands obtained for $c_0 = 0.01$, $P_{dES} = 0.01$, $P_{iES} = 0.0$, $p_w = 1.0$, $l_0 = 100$ and (a) $n_{DS} = 1$, time steps $10^5$, (b) $n_{DS} = 1$, time steps $5 \times 10^5$, and (c) $n_{DS} = 30$, time steps $10^5$. System size $100 \times 100$. Blue color denotes the first layer on flat surface.

The CA rules determine when an adatom is incorporated into the crystal. There are three different situations where an adatom becomes part of the crystal, as shown in Figure 1a. The usual places where an adatom attaches itself to a crystal are the kinks that are at the corners of the step. The second situation occurs when an adatom adjacent to a straight step, and at the same time, adjacent to another adatom becomes a crystal site. These rules determine the stiffness of the step. We assumed that the particles are easily built in the crystal at kinks, and more difficult at the straight part of the step. The step stiffness can be regulated, making a second event to be more or less likely. A step is stiffer when it is harder to build an adatom into the straight step. There is a third situation where the adatom turns into part of the crystal layer—the adatom becomes the nucleus of a new layer regardless of the step position. We assume that this is the case when at least three adatoms stick together. We are also adding a "correction" rule—if a single site is surrounded by steps from each side, it is filled whether there is an adatom or not. More details about the model are described in ref 18.

Above rules describe step flow during crystal growth and lead to various patterns on the crystal surface. These patterns are dependent on adatom diffusion. All particles diffuse independently along a given potential energy landscape that depends on the step position. Energy potentials on the surface are not an artificial creation but result from interactions on kinks and steps and/or from stresses on the surface related to, among others, surface reconstruction. All jumps along the terraces, except for those in the immediate vicinity of the step, are performed with the same probability, which is equal to 1 after the equal choice of jump direction. One of the possible mechanisms of adatom diffusion is the existence of a direct (also called usual) Ehrlich–Schwoebel (dES) barrier at the top of the step.[19−22] This makes jumping across the step (down or up) difficult. We set the probability of such jump $P_{dES}$ equal to 1 in the absence of a barrier and 0 for an infinite barrier. In a similar way, the iES barrier located at the bottom of step[19−22] with jump probability $P_{iES}$ was set. The presence of the iES makes adatom attachment to a step easier from the terrace behind than that from the one in front. Effects of barrier at the step are observed and analyzed based on experimental data.[23−26]

To get a more realistic model, we assumed a different potential energy at the bottom of the step due to the interaction with the particles that make up the crystal steps. For this reason, we added the parameter $p_w$ that determines the energy of the adatom remaining at the bottom of the step. Such an adatom, if its energy is higher than that in other positions, jumps over a barrier more easily, while if its energy is lower, its jump is more difficult. The parameter $p_w$ changes from 0 (which means that the particle is locked at its position) to a lower value of $P_{dES}^{-1}$ or $P_{iES}^{-1}$. The adatom jumps out of the site at the bottom of the step with a probability of $p_w P_{dES}$ or $p_w P_{iES}$. This parameter makes jumps through barriers become asymmetrical but satisfy the detailed balance condition. The visualization of initial barriers used in the presented paper are shown in Figure 1b–e.

Another possible mechanism of adatom diffusion is the applied directional bias $\delta$. All adatoms can diffuse along the system, jumping to the right with a probability of $1/2 + \delta$ or to the left with a probability $1/2 - \delta$. The bias is related to the diffusion asymmetry of adatoms. $\delta$ determines the probability of the adatom jumping in a preferred direction and according to the following equation

$$\frac{1}{2} + \delta = \frac{e^{\beta\eta}}{e^{\beta\eta} + e^{-\beta\eta}} \tag{3}$$

it varies from −0.5 to 0.5. The sign determines the direction of the applied bias: $\delta < 0$ induces step-up drift while $\delta > 0$ induces step-down drift. $\eta$ in eq 3 denotes the forcing force.

## ■ RESULTS AND DISCUSSION

**Pattern Formation with Increasing Diffusion.** The first example in which we show how a change in the diffusion rate will affect the surface pattern is systems with a low growth rate of crystals on which islands are built. Adatoms could be built into the crystal in the kinks and form nuclei on the terraces, while the probability of adatoms sticking to a straight step was reduced to almost zero. It means that we allow for adatoms to be adjacent to the step in 1 out of 18 cases. In addition, we assumed an infinite iES barrier and a nearly infinite dES barrier given by the probability of jumping across the steps $P_{dES} = 0.01$ in Figure 1b. At room temperature $T = 300\ K$, assumption of $P_{dES} = 0.01$ means that the additional height of step barrier above usual barrier for diffusion is 0.13 eV. This is a low value that can be very easily fulfilled in real crystals by reconstruction at step as a result of which interactions are modified giving a little higher, new potential value. Such value of the Schwoebel barrier has been found in several crystals, refs 22, 25, and 26. Adoption of such growth conditions generally leads to the formation of fragmented islands on the surface of the crystal. In Figure 2, we show the results obtained after $10^5$ time steps, what at room temperature and assuming typical attempt frequency values $10^{13}$ and typical diffusion barrier 0.5 eV translates into time of 2 or 3 s. The shape of obtained islands are dendrite-like and in some sense resembles the experimentally observed structures in highly Si-doped GaN layers.[27] We observed that for higher values of diffusional rate islands grow larger with a more regular and dense structure. To be sure that the structure with lower $n_{DS}$ is not an early stage of structure with higher one, we performed calculations for longer time steps. Results are presented in Figure 2b. As one can notice, the structure is different than the one obtained for higher diffusion rate. The structure of island created for $n_{DS} = 1$ is more diffused, fractal-like while in Figure 2c we obtained islands of rather compact form. This example illustrates that the form of islands growing on a crystal surface strongly depends on the diffusion rate. Fractal structures are formed in diffusion-limited growth, while kinetic-limited growth leads to the formation of more compact islands.

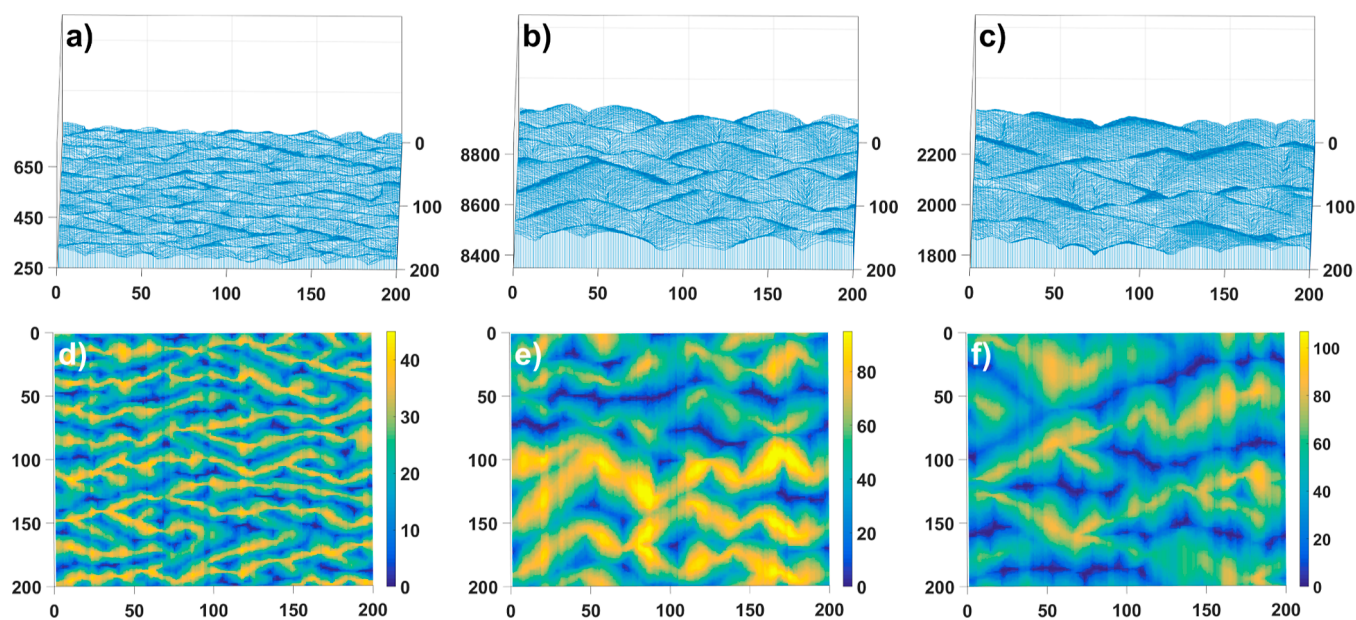

**Figure 3.** Side (top panel) and top (bottom panel) view of meanders obtained for $P_{dES} = 0.0$, $P_{iES} = 1.0$, $p_w = 1.0$, $l_0 = 5$ (a,d) $n_{DS} = 1$; (b,e) $n_{DS} = 5$; and (c,f) $n_{DS} = 10$; $c_0 = 0.02$ and time steps $10^7$. System size $200 \times 200$.

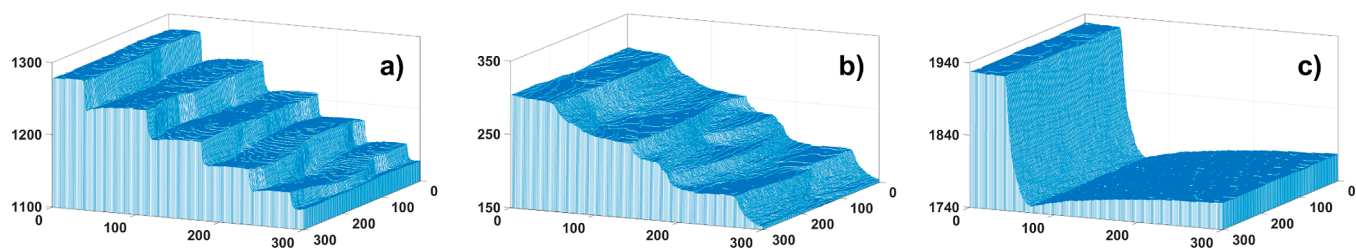

**Figure 4.** Bunches obtained for $P_{dES} = 1.0$, $P_{iES} = 1.0$, $p_w = 1.0$, bias = 0.1, $l_0 = 2$ and (a) $n_{DS} = 1$, time steps $5 \times 10^6$; (b) $n_{DS} = 30$, time steps $2 \times 10^5$; and (c) $n_{DS} = 30$, time steps $5 \times 10^6$; $c_0 = 0.02$. System size: $300 \times 300$.

To analyze the meanders, one of the regular patterns formed at the surface, we increased the probability of adatom to crystal incorporation at the straight step and simultaneously decreased it for nucleation. It means that we allow for nucleation only in one out of two cases and 6 out of 18 cases for adatoms adjacent to the step. In accordance with the literature[28−31] for this structures we have considered only the presence of an infinite dES barrier located at the top of the step with the probability of jumping across the steps $P_{dES} = 0$ (Figure 1c). We performed numerous simulation runs, increasing only the diffusion rate in each of them. It means that in our model, we increased the number of diffusional jumps $n_{DS}$. Results obtained after $10^7$ vicCA simulation time steps are presented in Figure 3 for three different values of $n_{DS}$: 1, 5, and 10. As one can notice, in the case of the shortest diffusion rate, we received several meanders with a short wavelength. With increasing the diffusion rate, we obtained longer meanders wavelength. For $n_{DS} = 1, 5, 10$, we can estimate wavelength as $\lambda = 16, 34$, and 55 interparticle distances. We can say that $\lambda \approx \sqrt{n_{DS}}$ what resembles the usual diffusion dependence between distance and time. Regardless, it is clear that the meandering process depends on the diffusion rate; thus, it can be controlled by change of the temperature.

Next case we want to investigate is the bunching process. This happens when infinite iES barrier is set with $P_{iES} = 0$.[14,16,28,32] The bunching takes place without any changes if we allow the particles to stick together, like in the first process of island grow. However, in the presence of only iES, the creation of bunches is slow over time. Therefore, in order to have faster and higher bunches, for which the diffusion rate effect would be more visible, we added step-down bias $\delta = 0.1$ to the system and at the same time resigned from the ES barriers, as shown in Figure 1c. Such choice of $\delta$ means that additional energy gained by a given particle is equal to 0.003 eV. In Figure 4, we present the obtained results. It can be seen that the structures retain initial, stepped shape due to the growth dynamics of the bunches, i.e., all the material is attached to the steps. More diffusional jumps, $n_{DS} = 30$, after the same simulation time steps ($5 \times 10^6$), lead to creation of macrostep. One can also notice that for structures with lower diffusion rate, the bunches are almost vertical, while for the one with higher diffusion rate, the bunches are stretched with gentler slope, in particular at the bottom of bunch. This effect is more visible when we compare structures not after the same simulation time steps but at the same stage of formation of the structures (see Figure 4a,b). Such a comparison also confirms that different diffusion rates do indeed lead to different structures. This observation is in agreement with results obtained in 1D case, where also higher diffusion rate leads to creation of macrosteps.[14] This should be related to the fact that when we go to the kinetic-limited growth mode, the diffusion becomes faster than the process of attaching of atoms to the step. The greatest effect resulting from the change in the rate of the diffusion process was obtained with the simultaneous presence of both the direct and iES barriers, as shown in Figure 1e. Nucleation of adatoms at the terrace is allowed if three of them meet together. The outcome after $10^6$ vicCA simulation time steps for $P_{dES}$ = 0.2, $P_{iES}$ = 0.4, and corresponding value $p_w$ = 2.5 can be seen in Figure 5. Such choice of parameters leads to creation of 3D structures, as shown in ref 18. At the same time, they mean very week perturbation of step potentials, namely, 0.05 eV at the top of step and 0.03 eV at the bottom, at room temperature 300 K. Such small modification of potential energy at steps are very likely in the real system through lattice reconstructions at steps.[23] However, in the case of the DL growth mode, we obtained a planar surface with different sizes of islands on it (Figure 5a). On the other side, namely, closer to the KL growth mode, very tall NWs appear (Figure 5c). Received structures are 5 orders of magnitude higher. In between, we found another type of pattern on the surface: broad nanopillars (Figure 5b). These structures are separated by deep and narrow cracks which in experiments were also observed.[33−36] Typically, in an experiment, the mechanism responsible for this is stress present in the structures or a thermal effect. Another possibility for the appearance of such cracks could also be surface instability resulting from surface deformation by the rearrangement of atoms. The dynamics of this process is usually assumed to be surface diffusion for a free surface.[37] Since we study the effect of the diffusion rate, this could be the reason cracks appear in our structures.

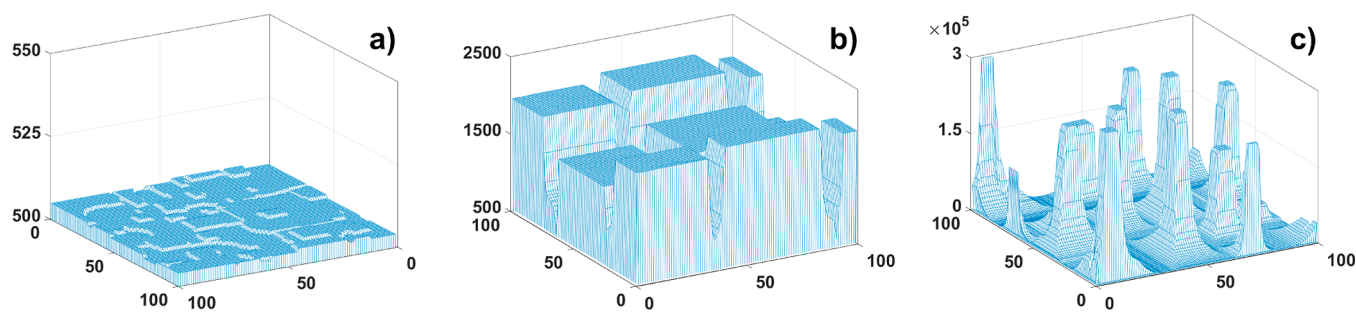

**Figure 5.** Structures obtained for $P_{dES} = 0.2$, $P_{iES} = 0.4$, $p_w = 2.5$, $l_0 = 100$ and (a) $n_{DS} = 5$, (b) $n_{DS} = 20$, and (c) $n_{DS} = 60$; $c_0 = 0.02$ and time steps $10^6$. System size $100 \times 100$. Note the difference in the vertical scale.

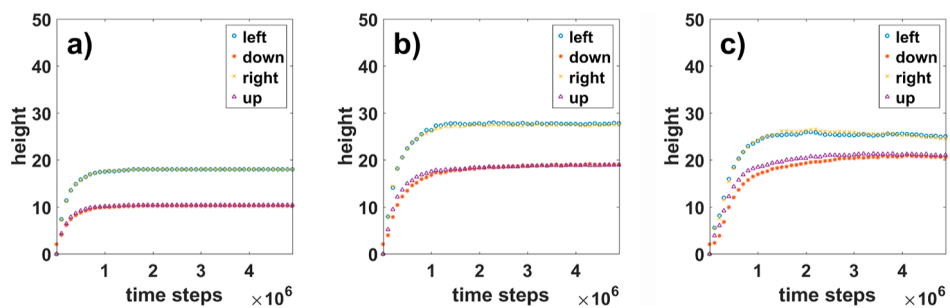

**Figure 6.** Height obtained for meanders. Presented results are for meanders with (a) $n_{DS} = 1$, (b) $n_{DS} = 5$, and (c) $n_{DS} = 10$.

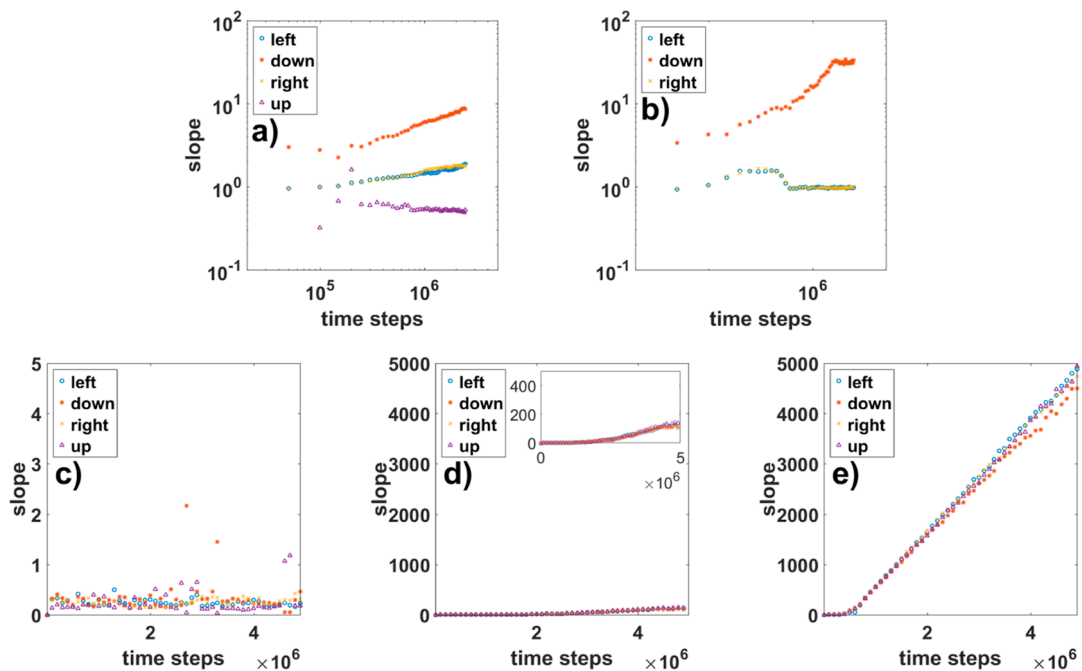

**Figure 7.** Slope obtained for bunches (middle panel), planar surface, nanopillars, and nanocolumns (bottom panel). Presented results are for bunches (in the logarithmic scale) with (a) $n_{DS} = 1$ and (b) $n_{DS} = 30$; for planar surface with (c) $n_{DS} = 5$, nanopillars with (d) $n_{DS} = 20$ and nanocolumns with (e) $n_{DS} = 60$. For better visibility of the slope in the case of nanopillars, we put the results in the inset with a 10 times smaller vertical scale.

We observe that as the diffusion rate increases, we obtain different domain shapes, larger bunches and meanders, and a change in the surface pattern from planar surface trough cracked nanopillars to tall NWs.

**Quantitative Measure of the Pattern Formation Process.** The variety of obtained structures makes it necessary to have an appropriate tool to identify patterns in a simple, automatic way. Therefore, we carried out a quantitative analysis by calculating different length scales: mean height and planar distance between the obtained structures as a function of the simulation time steps. The calculations were performed in four directions: down and up the steps and to the right and left along the steps. This directions are specified with respect to the initial surface arrangement. All results were averaged over 5 run.

Figure 6 shows how, for meanders created at different diffusion rates, the height of the structures thus calculated changes. According to the qualitative results presented in Figure 3, it can be seen that the mean height increases with increasing diffusion length in all four directions. From the difference between directions across and along the steps, we are able to recognize the structure expanded in one direction. In all cases, these quantities saturate after a certain time of simulation, but the saturation occurs at different levels for different diffusion rates. We also see that with a higher rate of diffusion, the structures began to have a comparable slope along and across the steps. This indicates a change in the nature of growth. With faster diffusion, no more flat meander structures are formed, but three-dimensional shapes more like pyramids. The results, presented as the ratio of the mean height to the planar distance, i.e., slope, are shown in Figure 7. In the case of the bunching process, the average slope does not saturate over time, it only increases. The results are shown in Figure 7a,b on a logarithmic scale. For bunches in the case of diffusion-limited growth, a large difference can be observed between the results calculated down the steps and along the steps (to the right and to the left). This shows a highly anisotropic structure in the direction perpendicular to the initial steps. The slope of these bunches scales over time as

$N(t) \sim t^{0.45}$. This is because the height of bunches grows with typical scaling parameter $\beta$ = 0.5 and the width of bunches scales as $1/z$ = 0.05.[16,17] The slope obtained up the steps mainly reflects the behavior of the single steps. At a higher diffusion rate, we still see high anisotropy in one direction; however, in this case, the bunches form later, but once they start, the process is fast. Due to the lack of structures, the up-step slope for higher diffusion rates is not shown. When both direct and iES barriers are present, the slope for each diffusion rate considered looks different and differs from the previous cases, as shown in Figure 7c−e. Here for the planar surface with islands, as expected, the resulting slope is almost constant (Figure 7c). When we look at the results received for $n_{DS}$ = 20 and 60, we notice that this quantity instantly grow at a different rate for each $n_{DS}$. In the case of NWs (Figure 7e), it scales almost linearly, while for nanopillars, it changes much slower but also increases monotonically. Each of the characteristics shown is different and can be used to identify the type of surface arrangement.

The measure we used to characterize the pattern formation process is related to the length of the correlation function and allows us to distinguish between various patterns without image analysis.

## ■ CONCLUSIONS

We have presented a qualitative and quantitative study of the dependence of the crystal surface pattern on the diffusion process using the cellular automata model. We showed how diffusion affects the growth dynamics and surface ordering. By changing the number of diffusion jumps, the transition from a diffusion-limited process to a kinetic-limited process was investigated. In the experimental situation, a larger diffusion parameter $n_{DS}$ means increased temperature. In the cases analyzed above, the temperature should increase by 50−80 K, which is not a big difference but causes serious changes in the process of surface pattern formation. The rate of diffusion along the terraces changes the shape of the growing islands from fragmented, fractal-like shapes for $n_{DS}$ = 1 to more compact structures when diffusion is faster. In the case of meanders, changes in the diffusion rate cause the wavelength of the meanders to increase or even create pyramidal structures for faster diffusion. Quantitatively, this process can be investigated by examining the time dependence of the heights of bunch-like structures in all direction. In the case of meanders, the level of saturation of these heights increases with the increase of $n_{DS}$, and the direction down-steps differs from the directions along the steps. During the bunching process, apart from the growth rate, the main difference resulting from the acceleration of the diffusion process is the change in the slope of the bunches. For slow diffusion, the bunches are almost vertical, while for higher ones, they are stretched out with a gentler slope, especially at the foot of the bunches. The detection of the average slopes of the bunches does not reflect changes in this kind of behavior, but shows their straightening with increased diffusion, as the slopes of the bunches decreases in the direction along the steps. The last analyzed example was devoted to growth with two types of ES barriers—direct and inverse. In this case, we observed the greatest effect of diffusion rate changes. For each diffusion rate presented, we obtained significantly different surface patterns. The slowest diffusion gave a flat surface. When the diffusion rate increased, going into a kinetic-limited growth mode, we switch between nanopillars with deep cracks to fast-growing, very tall NWs.

Characterization of structures by bunch slope makes it possible to distinguish between different surface patterns. All the results presented above illustrate the large influence of the diffusion rate on the nature of crystal growth. It follows that the diffusion rate in general can be used as a parameter to control the crystal growth mode.

## ■ AUTHOR INFORMATION

**Corresponding Author**

**Marta Anna Chabowska** − Institute of Physics Polish Academy of Sciences, 02-668 Warsaw, Poland; ● orcid.org/0000-0002-8500-3889; Email: galicka@ifpan.edu.pl

**Author**

**Magdalena A. Załuska-Kotur** − Institute of Physics Polish Academy of Sciences, 02-668 Warsaw, Poland; ● orcid.org/0000-0003-0488-8425

Complete contact information is available at:
https://pubs.acs.org/10.1021/acsomega.3c06377

**Notes**
The authors declare no competing financial interest.

## ■ ACKNOWLEDGMENTS

Part of the calculations were carried out on HPC facility Nestum (BG161PO003-1.2.05). Research is financially supported by The Polish National Center for Research and Development (grant no. EIG CONCERT-JAPAN/9/56/AtLv-AIGaN/2023).

## ■ REFERENCES

(1) Kang, J.-H.; Krizek, F.; Zaluska-Kotur, M.; Krogstrup, P.; Kacman, P.; Beidenkopf, H.; Shtrikman, H. Au-Assisted Substrate-Faceting for Inclined Nanowire Growth. *Nano Lett.* **2018**, *18* (7), 4115−4122.

(2) Arora, S. K.; O'Dowd, B. J.; Ballesteros, B.; Gambardella, P.; Shvets, I. V. Magnetic properties of planar nanowire arrays of Co fabricated on oxidized step-bunched silicon templates. *Nanotechnology* **2012**, *23*, 235702.

(3) Ortega, J. E.; Vasseur, G.; Piquero-Zulaica, I.; Matencio, S.; Valbuena, M. A.; Rault, J. E.; Schiller, F.; Corso, M.; Mugarza, A.; Lobo-Checa, J. Structure and electronic states of vicinal Ag(111) surfaces with densely kinked steps. *New J. Phys.* **2018**, *20* (7), 073010.

(4) Benoit−Maréchal, L.; Jabbour, M. E.; Triantafyllidis, N. Scaling laws for step bunching on vicinal surfaces: Role of the dynamical and chemical effects. *Phys. Rev. E* **2021**, *104*, 034802.

(5) Do, E.; Park, J. W.; Stetsovych, O.; Jelinek, P.; Yeom, H. W. Z3 Charge Density Wave of Silicon Atomic Chains on a Vicinal Silicon Surface. *ACS Nano* **2022**, *16*, 6598−6604.

(6) Gleißner, R.; Noei, H.; Chung, S.; Semione, G. D. L.; Beck, E. E.; Dippel, A.-Ch.; Gutowski, O.; Gizer, G.; Vonk, V.; Stierle, A. Copper Nanoparticles with High Index Facets on Basal and Vicinal ZnO Surfaces. *J. Phys. Chem. C* **2021**, *125*, 23561−23569.

(7) Ai, W.; Chen, X.; Feng, J. Microscopic origins of anisotropy for the epitaxial growth of 3C-SiC (0001) vicinal surface: A kinetic Monte Carlo study. *J. Appl. Phys.* **2022**, *131*, 125304.

(8) Krasteva, A.; Popova, H.; Krzyżewski, F.; Załuska-Kotur, M.; Tonchev, V. Unstable vicinal crystal growth from cellular automata. *AIP Conf. Proc.* **2016**, *1722*, 220014.

(9) Sudoh, K.; Iwasaki, H. Step dynamics in faceting on vicinal Si(113) surfaces. *J. Phys.: Condens. Matter* **2003**, *15*, S3241−S3253.

(10) Néel, N.; Maroutian, T.; Douillard, L.; Ernst, H.-J. Spontaneous structural pattern formation at the nanometre scale in kinetically restricted homoepitaxy on vicinal surfaces. *J. Phys.: Condens. Matter* **2003**, *15*, S3227−S3240.

(11) Rahman, T. S.; Kara, A.; Durukano lu, S. Structural relaxations, vibrational dynamics and thermodynamics of vicinal surfaces. *J. Phys.: Condens. Matter* **2003**, *15*, S3197−S3226.

(12) Minoda, H. Direct current heating effects on Si(111) vicinal surfaces. *J. Phys.: Condens. Matter* **2003**, *15*, S3255−S3280.

(13) Rousset, S.; Repain, V.; Baudot, G.; Garreau, Y.; Lecoeur, J. Self-ordering of Au(111) vicinal surfaces and application to nanostructure organized growth. *J. Phys.: Condens. Matter* **2003**, *15*, S3363−S3392.

(14) Krzyżewski, F.; Załuska-Kotur, M.; Krasteva, A.; Popova, H.; Tonchev, V. Step bunching and macrostep formation in 1D atomistic scale model of unstable vicinal crystal growth. *J. Cryst. Growth* **2017**, *474*, 135−139.

(15) Toktarbaiuly, O.; Usov, V. O.; Ó Coileáin, C.; Siewierska, K.; Krasnikov, S.; Norton, E.; Bozhko, S. I.; Semenov, V. N.; Chaika, A. N.; Murphy, B. E.; Lübben, O.; Krzyżewski, F.; Załuska-Kotur, M. A.; Krasteva, A.; Popova, H.; Tonchev, V.; Shvets, I. V. Step bunching with both directions of the current: Vicinal W(110) surfaces versus atomistic-scale model. *Phys. Rev. B* **2018**, *97*, 035436.

(16) Krzyżewski, F.; Załuska-Kotur, M.; Krasteva, A.; Popova, H.; Tonchev, V. Scaling and Dynamic Stability of Model Vicinal Surfaces. *Cryst. Growth Des.* **2019**, *19*, 821−831.

(17) Popova, H.; Krzyżewski, F.; Załuska-Kotur, M. A.; Tonchev, V. Quantifying the Effect of Step−Step Exclusion on Dynamically Unstable Vicinal Surfaces: Step Bunching without Macrostep Formation. *Cryst. Growth Des.* **2020**, *20*, 7246−7259.

(18) Załuska-Kotur, M.; Popova, H.; Tonchev, V. Step Bunches, Nanowires and Other Vicinal "Creatures"—Ehrlich−Schwoebel Effect by Cellular Automata. *Crystals* **2021**, *11*, 1135.

(19) Ehrlich, G.; Hudda, F. G. Atomic View of Surface Self-Diffusion: Tungsten on Tungsten. *J. Chem. Phys.* **1966**, *44*, 1039−1049.

(20) Schwoebel, R. L.; Shipsey, E. J. Step Motion on Crystal Surfaces. *J. Appl. Phys.* **1966**, *37*, 3682−3686.

(21) Pimpinelli, A.; Videcoq, A. Novel mechanism for the onset of morphological instabilities during chemical vapour epitaxial growth. *Surf. Sci.* **2000**, *44S*, L23−L28.

(22) Bellmann, K.; Pohl, U. W.; Kuhn, Ch.; Wernicke, T.; Kneissl, M. Controlling the morphology transition between step-flow growth and step-bunching growth. *J. Cryst. Growth* **2017**, *478*, 187−192.

(23) Gerlach, R.; Maroutian, T.; Douillard, L.; Martinotti, D.; Ernst, H.-J. A novel method to determine the Ehrlich−Schwoebel barrier. *Surf. Sci.* **2001**, *480*, 97−102.

(24) Li, S.-C.; Han, Y.; Jia, J.-F.; Xue, Q.-K.; Liu, F. Determination of the Ehrlich-Schwoebel barrier in epitaxial growth of thin films. *Phys. Rev. B* **2006**, *74*, 195428.

(25) Gianfrancesco, A. G.; Tselev, A.; Baddorf, A. P.; Kalinin, S. V.; Vasudevan, R. K. The Ehrlich-Schwoebel barrier on an oxide surface: a combined Monte-Carlo and in situ scanning tunneling microscopy approach. *Nanotechnology* **2015**, *26*, 455705.

(26) Myint, P.; Erb, D.; Zhang, X.; Wiegart, L.; Zhang, Y.; Fluerasu, A.; Headrick, R. L.; Facsko, S.; Ludwig, K. F., Jr Measurement of Ehrlich-Schwoebel barrier contribution to the self-organized formation of ordered surface patterns on Ge(001). *Phys. Rev. B* **2020**, *102*, No. 201404(R).

(27) Sawicka, M.; Turski, H.; Sobczak, K.; Feduniewicz-Żmuda, A.; Fiuczek, N.; Gołyga, O.; Siekacz, M.; Muziol, G.; Nowak, G.; Smalc-Koziorowska, J.; Skierbiszewski, C. Nanostars in Highly Si-Doped GaN. *Cryst. Growth Des.* **2023**, *23*, 5093−5101.

(28) Misbah, C.; Pierre-Louis, O.; Saito, Y. Crystal surfaces in and out of equilibrium: A modern view. *Rev. Mod. Phys.* **2010**, *82*, 981−1040.

(29) Krzyżewski, F.; Załuska−Kotur, M. A. Coexistence of bunching and meandering instability in simulated growth of 4H-SiC(0001) surface. *J. Appl. Phys.* **2014**, *115*, 213517.

(30) Krzyżewski, F.; Załuska-Kotur, M. A. Stability diagrams for the surface patterns of GaN(000−1) as a function of Schoebel barrier height. *J. Cryst. Growth* **2017**, *457*, 80−84.

(31) Turski, H.; Krzyżewski, F.; Feduniewicz-Żmuda, A.; Wolny, P.; Siekacz, M.; Muziol, G.; Cheze, C.; Nowakowski-Szukudlarek, K.; Xing, H. G.; Jena, D.; Załuska-Kotur, M.; Skierbiszewski, C. Unusual step meandering due to Ehrlich-Schwoebel barrier in GaN epitaxy on the N-polar surface. *Appl. Surf. Sci.* **2019**, *484*, 771−780.

(32) Sato, M.; Uwaha, M. Growth law of step bunches induced by the Ehrlich−Schwoebel effect in growth. *Surf. Sci.* **2001**, *493*, 494−498.

(33) Hayafuji, N.; Kizuki, H.; Miyashita, M.; Kadoiwa, K.; Nishimura, T.; Ogasawara, N.; Kumabe, H.; Tada, T. M.; Tada, A. Crack Propagation and Mechanical Fracture in GaAs-on-Si. *Jpn. J. Appl. Phys.* **1991**, *30*, 459.

(34) Park, J.-S.; Tang, M.; Chen, S.; Liu, H. Heteroepitaxial Growth of III-V Semiconductors on Silicon. *Crystals* **2020**, *10*, 1163. and the references therein

(35) Wang, B.; Syaranamual, G. J.; Lee, K. H.; Bao, S.; Wang, Y.; Lee, K. E. K.; Fitzgerald, E. A.; Pennycook, S. J.; Gradecak, S.; Michel, J. Effectiveness of InGaAs/GaAs superlattice dislocation filter layers epitaxially grown on 200 mm Si wafers with and without Ge buffers. *Semicond. Sci. Technol.* **2020**, *35*, 095036.

(36) Sushkov, A.; Pavlov, D.; Andrianov, A. I.; Shengurov, V. G.; Denisov, S. A.; Chalkov, V. Yu.; Kriukov, R. N.; Baidus, N. V.; Yurasov, D. V.; Rykov, A. V. Comparison of III−V Heterostructures Grown on Ge/Si, Ge/SOI, and GaAs. *Semiconductors* **2022**, *56*, 122−133.

(37) Spatschek, R.; Brener, E.; Karma, A. Phase field modeling of crack propagation. *Philos. Mag.* **2011**, *91*, 75−95.